\documentclass{PoS}

\title{Modern middleware for the data acquisition of the Cherenkov Telescope Array}

\ShortTitle{Data Acquisition of the Cherenkov Telescope Array}

\author{Etienne Lyard\textsuperscript{$a$}, Roland Walter\textsuperscript{$a$}, Karl Kosack\textsuperscript{$b$}, Jean
Jacquemier\textsuperscript{$c$}, Igor Oya\textsuperscript{$d$}, Peter Wegner\textsuperscript{$d$}, Matthias
Fuessling\textsuperscript{$d$} and \speaker{Xin Wu}\textsuperscript{$a$} for the CTA Consortium \\
		\llap{\textsuperscript{$a$}}Observatory of Geneva, University of Geneva, Chemin d'Ecogia 19, CH-1290 Versoix, Switzerland;\hphantom{s}
		\llap{\textsuperscript{$b$}}CEA Saclay, DSM/IRFU/SAp, Bat 709, F-91191 Gif-Sur-Yvette, France;\hphantom{s}
		\llap{\textsuperscript{$c$}}Laboratoire d'Annecy-le-Vieux de Physique des Particules, Universit\'{e} de Savoie, CNRS/IN2P3, F-74941 Annecy-le-Vieux, France;\hphantom{s} 
		\llap{\textsuperscript{$d$}}DESY Zeuthen, Platanenallee 6, D-15738 Zeuthen, Germany\\
        E-mail: \email{etienne.lyard@unige.ch}, 
				\email{roland.walter@unige.ch}}

\abstract
{
The data acquisition system (DAQ) of the future Cherenkov Telescope Array (CTA)
 must be efficient, modular and robust to be able to cope with the very large 
data rate of up to 550 Gbps coming from many telescopes with different characteristics. 
The use of modern middleware, namely \texttt{ZeroMQ} and \texttt{Protocol Buffers}, can help to achieve 
these goals while keeping the development effort to a reasonable level. \texttt{Protocol 
Buffers} are used as an on-line data format, while \texttt{ZeroMQ} is employed to communicate 
between processes. The DAQ will be controlled and monitored by the Alma Common
Software (ACS).

\texttt{Protocol Buffers} from Google are a way to define high-level data structures 
through an interface description language (IDL) and a meta-compiler. \texttt{ZeroMQ} is a middleware 
that augments the capabilities of TCP/IP sockets. It does not implement very high-level 
features like those found in \texttt{CORBA} for example, but makes use of sockets easier, more robust and 
almost as effective as raw TCP. The use of these two middlewares enabled us to 
rapidly develop a robust prototype of the DAQ including data persistence to 
compressed FITS files.
}

\FullConference{The 34th International Cosmic Ray Conference,\\
		30 July- 6 August, 2015\\
		The Hague, The Netherlands}

\begin{document}

\section{Introduction}

\begin{figure}[h]
\centering     
\includegraphics[width=0.85\textwidth]{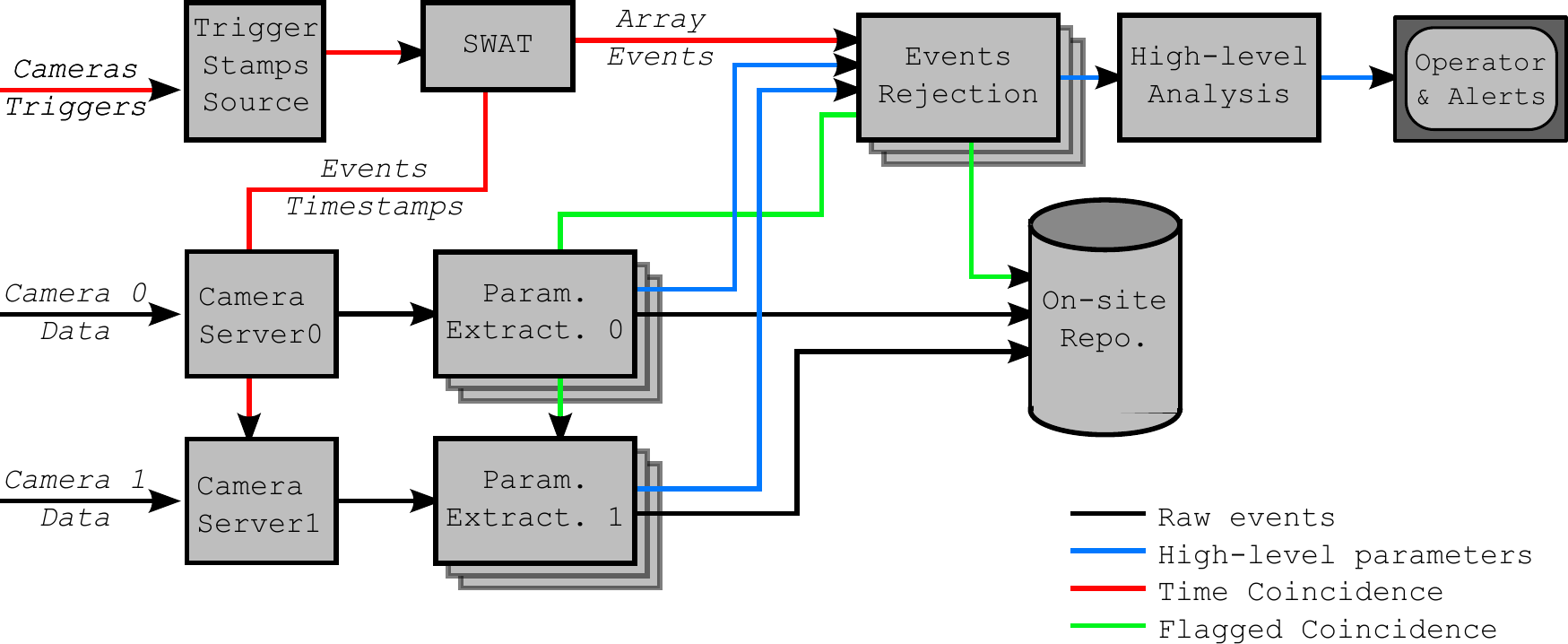}
\caption{ \label{fig:system_design}
Data flow diagram of the proposed real-time data reduction configuration. Single-camera triggers are time-stamped before being
passed to the software array trigger (SWAT). Timestamps passing the SWAT criteria are sent to the camera servers, which
in turn send the associated event data to the parameter extraction nodes. There the data are calibrated, a first analysis
performed and image parameters extracted. From there, event parameters are delivered to the event rejection module (blue
lines) where they are assembled into array-events and further filtered. The result of this analysis is fed back to the parameter
extraction nodes (green lines) and forward to the high-level analysis. Events that remain of interest are then compressed and
written to the repository (black lines), while the result of the high-level analysis is presented to the operator. The parameter
extraction nodes must provide enough buffering capabilities to wait until the analysis is complete before deciding whether to
discard an event or not. All boxes and connecting lines in this diagram are logical entities. More than one box can run on a
single physical server, and a given box can be load-distributed across several servers. The lines also represent logical connections,
which will be implemented via the \texttt{ZeroMQ} middleware for the data transfer and ACS remote calls for the configuration. In
order to improve fault tolerance, information from the event rejection could be routed in parallel to an algorithm that detects
and corrects faults in the information used to associate events. The results of that algorithm would then be provided to the
off-line analysis.}
\end{figure}

The Cherenkov Telescope Array \cite{cta} will generate large amounts of raw data. It will consist of a heterogeneous system 
composed of telescopes of different sizes, designs and cameras. The total throughput from the telescopes
is thought to be as high as 550 Gbps while this data is to be filtered so that less than 50 Gbps are stored on-site.  
Additionally, ways to further reduce the amount of raw data by another factor 10 are under investigation \cite{rta}.
To handle such large data rates along with the multiple filtering processes it is desirable that the 
Array Control software \cite{actl} and more specifically the data acquisition system (DAQ) is efficient, modular and robust. An illustration of a possible pipeline 
is illustrated in Fig. \ref{fig:system_design}. This architecture will be deployed on top of the Alma Common Software framework (ACS) \cite{acs} so that 
the DAQ can be controlled and monitored like any other component of the array. We propose a
modular architecture that brings several improvements compared to  a monolithic approach. Complex processing
 can be decomposed into simple tasks, thus easing the development and validation procedures. It allows for 
easier load-balancing across several physical nodes and helps to achieve the high availability requirement by 
allowing the reconfiguration of the system at any time. In addition to air-shower observations, it is proposed
that CTA will also operate in optical interferometry mode \cite{interferometry}. This extra functionality
would continuously record the observed value of a single pixel, sampled at 100 MHz or more. Therefore 
the DAQ is designed to have the necessary level of modularity, which allows for observing mode
reconfiguration on a short timescale.

The DAQ will make use of three main external libraries. 
The \texttt{Protocol Buffers} library from Google \cite{protobufs}, is used for serializing structured
data. It maintains backward and forward compatibility and provides reflection.
\texttt{ZeroMQ} \cite{zmq} is a middleware that simplifies the usage of streams. Although designed for
ALMA \cite{alma}, it is proposed that the ACS framework will be used to deploy, control and monitor
all software components of CTA.

\section{Data Volume Reduction Scenarios}
The general concept for camera data capture involves transmission from each camera to a camera
server process running at a central location. In most cases a stream of all available raw data (i.e.
locally triggered events) can be continuously delivered, without any event pre-selection (or array-level
triggering). This approach minimises the complexity on the camera side and delivers maximum flexibility
for data collection. Subsystem-level hardware triggers are envisaged internally by the camera groups
of the large-sized and Schwarzschild-Couder telescopes. Under these assumptions, the volume of data 
delivered by cameras must be reduced by a large factor ($\approx 100$) so that it can be written to permanent storage.
Reducing this large data set should not discard scientifically useful data, thus each new data
trimming process should be carefully evaluated and tested off-line \cite{pipelines} before it is introduced in the real-time
analysis pipeline \cite{rta}. The proposed modular architecture achieves this by storing the full dataset during
commissioning and gradually introducing new data trimming algorithms in the real-time pipeline once
they are well understood.

Various data reduction strategies are under consideration, including obvious choices such as \emph{waveform
reduction} and \emph{zero suppression}, but also more ambitious ones such as \emph{background suppression}. The
latter aims at discarding proton and helium nuclei initiated cascades, from which most of the background Cherenkov light is produced.
Tagging such events in a reliable way is a computationally intensive process, which will be challenging
to implement in real-time. Further details on these approaches, as well as the implication on the data
pipelines and calibration are provided in \cite{datareduction}.

\section{Data Streams}

\begin{figure}[h]
\centering     
\includegraphics[width=0.85\textwidth]{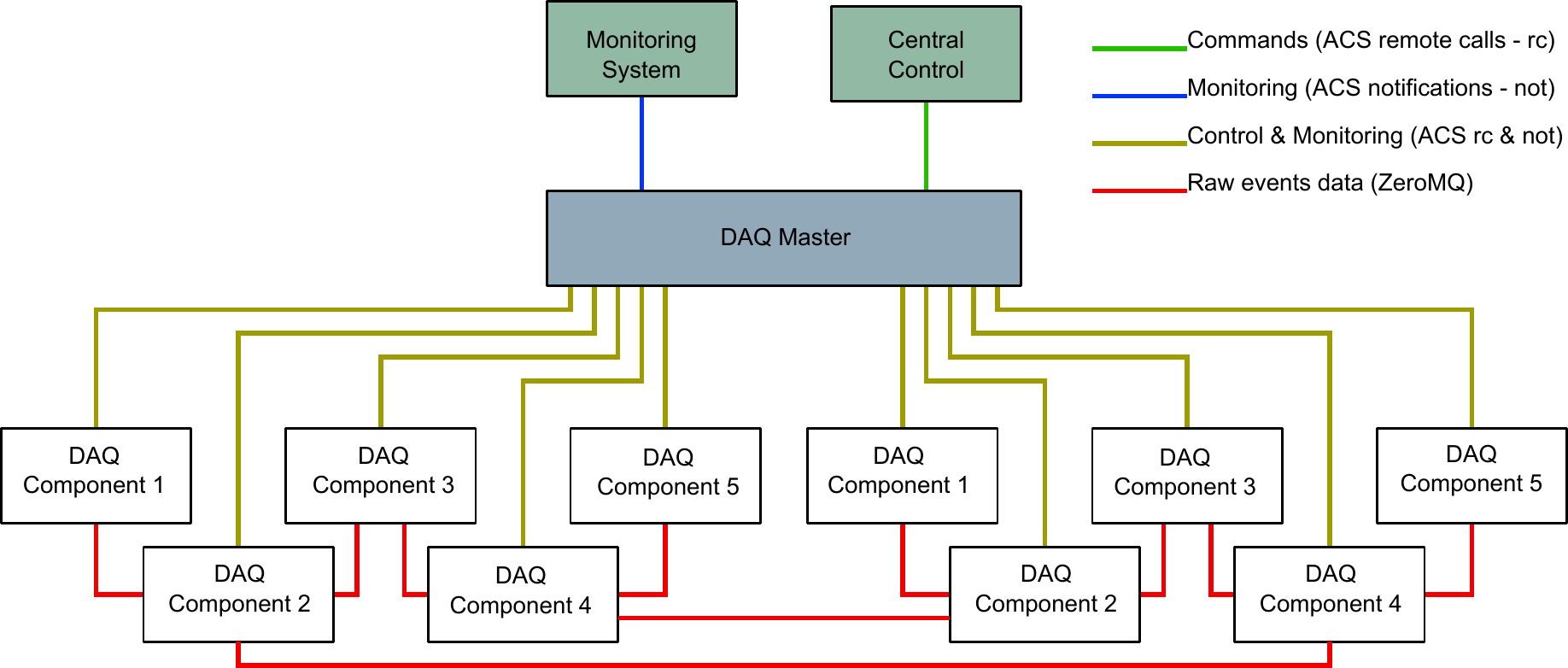}
\caption{ \label{fig:ACS}
Conceptual view of the DAQ data flow with an arbitrary number of low-level components inter-connection. 
Commands are provided by the central control, directly to the DAQ master. The master then takes appropriate action
to reconfigure the relevant components accordingly, for instance by connecting two low-level components with a raw
events stream. Each low-level component is monitored by the DAQ master which in turn forwards summary information 
to the monitoring system. If necessary, such as for debugging purposes, it is possible that the monitoring
system directly connects to the low-level components notification channels. The vertical data flow is entirely handled
by ACS (either through remote procedure calls or notification channels) while the horizontal, high throughput data
flow is entirely handled by \texttt{ZeroMQ} streams. This architecture reduces the complexity for the central control, and lets
the DAQ master handle errors.}
\end{figure}

All control and monitoring will be implemented using ACS, while the bulk data transfer will be
implemented using \texttt{ZeroMQ}, as seen in figure \ref{fig:ACS}. 
All connection procedures and the messages concept will be implemented with \texttt{ZeroMQ}. 
Because \texttt{ZeroMQ} is also capable of connecting different processes
or threads running on a single machine, it is very easy to build single large components
from several smaller ones if the communication performance requirements are not met by network protocols.
Connection paradigms will be chosen individually for each connection item. The possible cases
are request-reply, push-pull and publish-subscribe. 
\texttt{ZeroMQ} implement all connection procedures along with the concept of messages. This way the user does not
need to worry about buffers or protocols. \texttt{ZeroMQ} delivers incoming messages as they arrive, depending on the 
selected distribution paradigm. 

\section{Structured Data Format}
The internal DAQ format will rely on \texttt{Protocol Buffers} developed and widely used by Google \cite{protobufs}.
\texttt{Protocol Buffers} provide a standardised way to handle the serialization problem. Structures are
defined via a high-level interface description language. A compiler then
generates the run-time code. The serialized version of the structure is highly optimized, with values
being packed as efficiently as possible to reduce the network throughput or left untouched to minimize
the serialization overload. Moreover, the built-in forward and backward compatibility of the format allows
modification of the structure content without breaking the pipeline. As an option that can be selected,
the built-in reflection of the structures allows reading and writing of any data structure to/from persistent
storage without having to write a single line of code specific to that data structure.

Last but not least, \texttt{Protocol Buffers} are available in many programming languages. This will allow us to mix different
languages to build the real-time analysis pipeline and thus to reuse offline algorithms that may have been
written in languages other than C/C++. 

One weakness of \texttt{Protocol Buffers} is that they do not include a 16-bit integer type. We
overcame this problem by implementing our own array class that operates on top of the standard arrays 
in protocol buffers. In addition, the serialisation is sometimes slower than expected. To remedy this,
we had to use the fixed32 type instead of the int32 type. This prevented the occurrence of time-consuming
on-the-fly compression. 

\section{Prototyping}

As a proof of concept, a prototype involving the components described below was developed:

\textbf{Data Buffers} store events and/or event parameters in memory for some time until a
decision is taken regarding their fate. They provide flexibility in the pipeline by allowing
the array event builders to retrieve event parameters only when the time coincidence is known.

\textbf{Repository Writers} are responsible for committing incoming data to the on-site
repository. If the data is provided raw then it is compressed on-the-fly. Each repository
writer deals with one single stream and opens new files once the size or event number
limit is reached or after a new run is started. The target storage node will remain the same at least
during a single run. Changing the storage node from run to run would allow for the next-day analysis to
start while data taking is still ongoing, possibly providing more accurate results earlier.

\textbf{Processing Nodes} are meant to be all-purpose skeleton components so that any kind of computation
can easily be introduced to the pipeline. Their primary goal is to embed real-time processing
into the DAQ pipeline. For instance, extracting the image parameters from the raw event data is 
desirable early in the pipeline to help reduce the total network throughput. The data reduction
algorithms will be provided by another workpackage (DATA). We developed a simple image cleaner and muon extractor based 
on the Hough transform \cite{hough} to use within this demonstrator.
 
\begin{figure}[h]
\centering     
\includegraphics[width=0.7\textwidth]{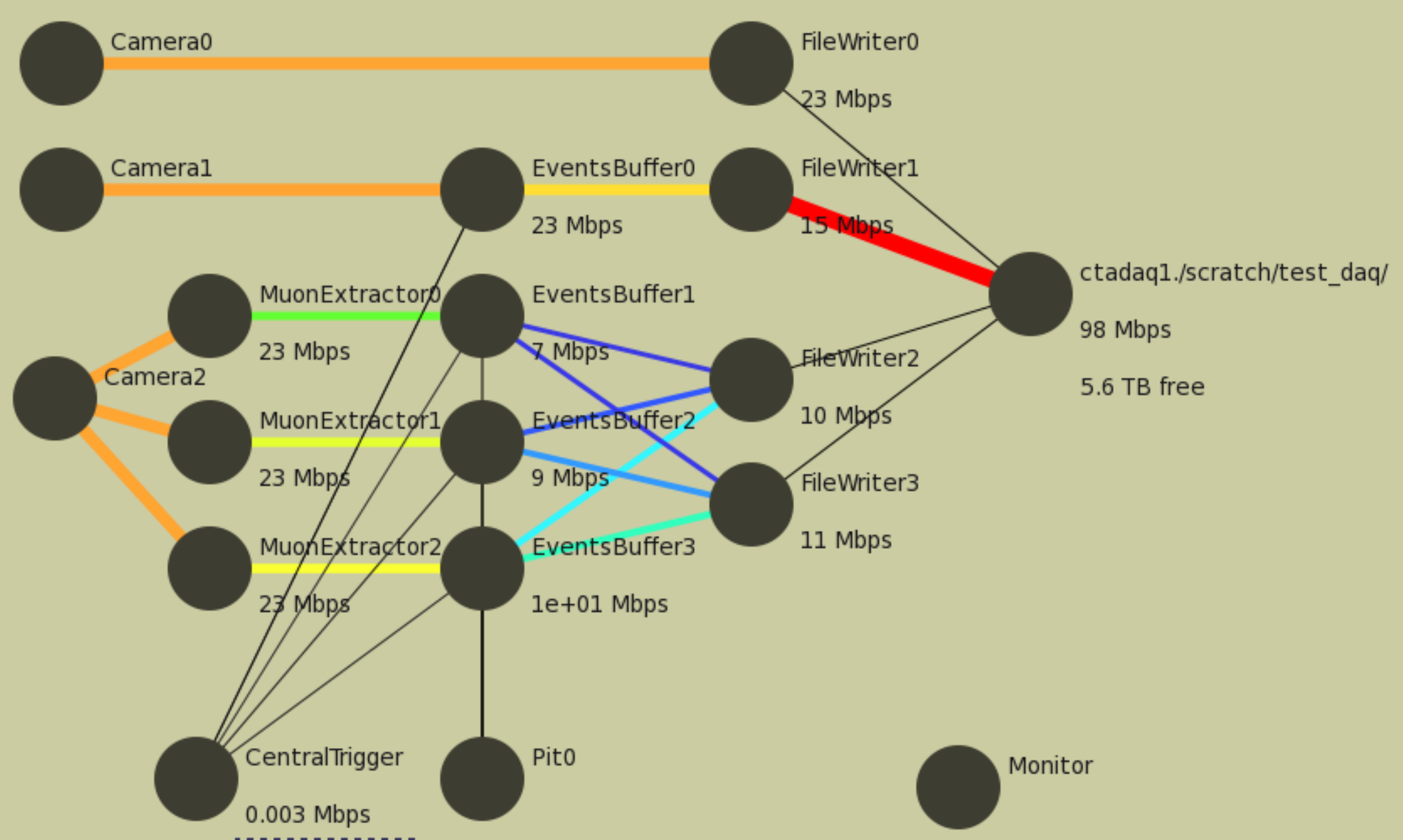}
\caption{ \label{fig:pipeline}
Screenshot of the monitor display from the DAQ demonstrator. Each circle represents a DAQ component
while the lines connecting them are actual \texttt{ZeroMQ} connections. The colour and width of the lines reflect the average throughput for
each connection. Due to the use of \texttt{ZeroMQ} and \texttt{Protocol Buffers}, new components can be added or removed without
stopping operations. This particular view shows the data taking process from 3 cameras. The first one has all its events written to
disk, the second camera pipeline discards events based on their stereo coincidence while the third camera pipeline has an
image cleaner and muon extractor which analyses the data in real-time. The EventsBuffer nodes retain the events until their
stereo coincidence is known, while the FileWriter nodes compress and write the events to disk in compressed FITS format. The
node called \emph{Pit0} is a proxy that only receives events classified as muons for monitoring purposes
}
\end{figure}

We also developed a monitoring tool to be used in place of ACS so that the required third party software is
reduced to a minimum. A display of the running pipeline can be seen in figure \ref{fig:pipeline}. Due to
time constraints, this prototype only focuses on load-balancing and automatic failure recovery. The maximum
throughput was evaluated between two nodes only and reached $\approx 9$ Gbps. This suggests that the chosen
middleware makes the pipeline perform $\approx 10 \%$ slower compared to when using raw C-structs and tcp sockets
in low-latency environments. This should not be an issue as more than one stream can easily be 
created to accommodate higher throughput.  

Integration tests to verify if this approach is suitable to readout the data from the camera
servers is yet to be implemented. In addition, these tests will highlight if any optimizations
and/or specific constraints are required. The small-sized telescopes should not present a problem, as the data rates remain low. 
For the larger-sized telescopes the high throughput from the cameras will most likely prevent the DAQ 
from performing on-the-fly data conversion. Thus delaying the unification of the data format to a later 
stage in the acquisition pipeline.

The prototype pipeline allows for nodes to be discarded or new ones to be introduced at any time. 
For example, if a server fails and all data sitting in its memory is lost but all other data manages to 
buffer (until a replacement becomes available). 

Using the \texttt{Protocol Buffers} allowed us to make our data model evolve in a flexible way. This evolution was triggered by the introduction of new 
concepts that required new data structures or the modification of existing ones. 

\section{Conclusion}

The use of the proposed middlewares allowed us to quickly develop 
a proof-of-concept toy  model. This would not have been possible using traditional sockets and structures as many
low-level features would have had to be reimplemented. Using these middlewares to achieve greater flexibility
comes at a computing power and throughput cost of $\approx 10 \%$. We strongly believe that this
approach is the best one to follow because high-level implementations can be replaced by lower-level code
that is optimised toward the chosen features and architecture of the system, and which is unlikely to 
change over time.


\begin{thebibliography}{99}

\bibitem{cta}
B.S. Acharya et al.,
\emph{Introducing the CTA concept},
\emph{Astroparticle Physics} {vol. 43 No. 3}, 2013

\bibitem{rta}
A. Bulgarelli, V. Fioretti, 
\emph{The On-Site Analysis of the Cherenkov Telescope Array},
In these proceedings, 2015

\bibitem{actl}
I. Oya et al., 
\emph{Status and Plans for the Array Control and Data Acquisition system of the Cherenkov Telescope Array}
In these proceedings, 2015

\bibitem{acs}
G. Chiozzi et al.,
\emph{The ALMA common software: a developer-friendly CORBA-based framework.}
in Advanced Software, Control, and Communication Systems for Astronomy, 
Society of Photo-Optical Instrumentation Engineers (SPIE) Conference Series 5496, 205-218, 2004

\bibitem{interferometry}
D. Dravins et al.,
\emph{Optical intensity interferometry with the Cherenkov Telescope Array.}
\emph{Astroparticle Physics} {vol. 43 No. 0}, 2013

\bibitem{protobufs}
Google Inc.,
\emph{Protocol Buffers},
http://developers.google.com/protocol-buffers, 2014

\bibitem{zmq}
Hintjens, P.,
\emph{ZeroMQ: Code Connected},
http://zeromq.org, 2014

\bibitem{alma}
A. Wootten,
\emph{The Atacama large millimeter/submillimeter array.},
in proceedings of the IEEE 97.8 : 1463-1471, 2009

\bibitem{pipelines}
K. Kosack et al.,
\emph{Cherenkov Telescope Array Data Processing Pipeline},
In these proceedings, 2015

\bibitem{datareduction}
J. Hinton et al., 
\emph{Data Volume Reduction in CTA},
in the CTA documentation, ref. MAN-PO/130828, 2014

\bibitem{hough}
P. E. Hart, 
\emph{How the Hough Transform Was Invented},
in IEEE Signal Processing Magazine, 26(6):18 - 22, 2009

\end{thebibliography}
\end{document}